\documentclass[12pt]{amsart}
\usepackage{amsmath,amssymb,cite,enumerate,graphicx,hyperref}

\hypersetup{pdftex,colorlinks=true,allcolors=blue}
\usepackage[all]{hypcap}

\usepackage[foot]{amsaddr}
\usepackage[sc]{mathpazo}

\newcommand\ack{\subsection*{Acknowledgment}}

\DeclareMathAlphabet\mathsfbi{T1}{phv}{b}{it}

\numberwithin{equation}{section}

\newcommand\BV{\boldsymbol} 
\newcommand\BM{\mathsfbi} 

\newcommand\dif{\mathrm{d}}
\newcommand\eqdef{\stackrel{\text{def}}=}
\newcommand\parderiv[2]{\frac{\partial #1}{\partial #2}}
\newcommand\deriv[2]{\frac{\dif #1}{\dif #2}}
\newcommand\expec{\mathbb E}
\newcommand\RN{{\mathbb R^N}}
\newcommand\MM{\mathcal M}
\newcommand\Lkolmsoln{\mathcal P}

\begin{document}

\title[Leading order response of statistical averages to stochastic
  perturbations]{Leading order response of statistical averages of a
  dynamical system to small stochastic perturbations}

\author[Rafail V. Abramov]{Rafail V. Abramov}

\date\today

\address{Department of Mathematics, Statistics and Computer Science,
University of Illinois at Chicago, 851 S. Morgan st., Chicago, IL 60607}

\email{abramov@math.uic.edu}

\begin{abstract}
The classical fluctuation-dissipation theorem predicts the average
response of a dynamical system to an external deterministic
perturbation via time-lagged statistical correlation functions of the
corresponding unperturbed system. In this work we develop a
fluctuation-response theory and test a computational framework for the
leading order response of statistical averages of a deterministic or
stochastic dynamical system to an external stochastic perturbation. In
the case of a stochastic unperturbed dynamical system, we compute the
leading order fluctuation-response formulas for two different cases:
when the existing stochastic term is perturbed, and when a new,
statistically independent, stochastic perturbation is introduced. We
numerically investigate the effectiveness of the new response formulas
for an appropriately rescaled Lorenz 96 system, in both the
deterministic and stochastic unperturbed dynamical regimes.
\end{abstract}

\maketitle

\section{Introduction}
\label{sec:intro}

Under suitable conditions, the fluctuation-dissipation theorem (FDT)
\cite{Kub1,Kub2,Ris} furnishes an approximation to the statistical
response of a dynamical system to a deterministic external
perturbation via statistical correlations of the unperturbed
dynamics. The FDT offers more insight into statistical properties of
dynamical processes near equilibrium in various scientific
applications
\cite{EvaMor,KubTodHas,Bel,Lei,CarFalIsoPurVul,GriBra,Gri,GriBraDym,GriBraMaj,GriDym,MajAbrGro,CohCra,MajAbrGer}. In
the past works
\cite{AbrMaj5,AbrMaj4,AbrMaj6,Abr5,Abr6,Abr7,Abr11,Abr12}, a
computational framework predicting the average response of both
deterministic and stochastic dynamical systems to a small
deterministic external perturbation was developed and extensively
studied.

In the current work, we develop the fluctuation-response theory and
numerically test the computational framework of the response of
statistical averages to a stochastic external perturbation, for both
the deterministic and stochastic unperturbed dynamics.  For the
deterministic unperturbed dynamics, our set-up is similar to the one
used in~\cite{Luc}, however, the resulting formula we arrive at is
different from the one in~\cite{Luc}, on which we will comment below.
For the stochastic unperturbed dynamics, we consider two different
types of perturbations: first, where the existing stochastic term is
perturbed, and, second, when a new, statistically independent,
stochastic term is introduced. We test the computational framework of
the stochastic response on the Lorenz 96 system~\cite{Lor,LorEma},
which we used as a test-bed nonlinear chaotic system with forcing and
dissipation for various purposes in the
past~\cite{AbrMaj4,AbrMaj5,Abr5,Abr6,Abr7,Abr8,Abr9,Abr10,Abr11,Abr12,AbrKje,MajAbrGro}.

Before going into the details, here we start by explaining the basic
idea of the average response and how it can be expressed via the
statistical properties of the underlying unperturbed dynamical system,
and also what problems one runs into while considering what otherwise
seems to be a rather simple dynamical set-up.

\subsection{Deterministic dynamics}

We start by considering a system of ordinary differential equations of
the form
\begin{equation}
\label{eq:dyn_sys}
\deriv{\BV x_t}t=\BV f(\BV x_t),
\end{equation}
where~$t$ is a scalar time variable, $\BV x_t$ is an $N$-dimensional
vector in the Euclidean space~$\RN$, and $\BV f:\RN\to\RN$ is a smooth
vector field. Observe that the solution $\BV x_t$ can be specified in
the form of a semigroup~$\phi_t$,
\begin{equation}
\label{eq:dyn_sys_soln}
\BV x_t=\phi_t\BV x=\BV x+\int_0^t\BV f(\BV x_s)\dif s,
\end{equation}
where $\BV x$ is the initial condition. We assume that any
solution~$\BV x_t=\phi_t\BV x$ of~\eqref{eq:dyn_sys} is attracted, as
$t\to\infty$, to a compact set~$\MM\subset\RN$, on which it possesses
a unique invariant ergodic measure~$\mu$. We will say that $\MM$ is
the global attractor of~\eqref{eq:dyn_sys}. Here we assume that the
system in~\eqref{eq:dyn_sys} is chaotic and mixing, that is, it has
positive first Lyapunov exponent and decaying time autocorrelation
functions.

Let~$A(\BV x)$ be a twice-differentiable function on~$\RN$, then we
denote its $\mu$-average as
\begin{equation}
\langle A\rangle=\int_\MM A(\BV x)\dif\mu(\BV x).
\end{equation}
Observe that even though~$A(\BV x_t)$ changes with time $t$, its
$\mu$-average $\langle A(\BV x_t)\rangle$ is fixed in $t$, due to the
fact that $\mu$ is invariant on $\MM$ under $\phi_t$,
\begin{equation}
\langle A(\BV x_t)\rangle=\int_\MM A(\phi_t\BV x)\dif\mu(\BV x)=
\int_\MM A(\BV x)\dif\mu(\BV x)=\langle A\rangle.
\end{equation}

\subsection{The concept of the average response to a stochastic
perturbation}

Consider the situation where the average $\langle A\rangle$ is
computed across a statistical ensemble of solutions
of~\eqref{eq:dyn_sys}, which is distributed according to the invariant
measure $\mu$ above.  As we already pointed out, this average $\langle
A\rangle$ is constant in time. However, assume that, at $t=0$, an
external perturbation (that is, a stochastic modification of the
right-hand side) is introduced into~\eqref{eq:dyn_sys}.  Clearly,
since the right-hand side is different, the measure $\mu$ is no longer
invariant for the new, modified dynamics. Because of that, the
statistical average $\langle A\rangle$ with respect to $\mu$ becomes
time-dependent for the perturbed dynamics. Here we do not assume that
$\mu$ is necessarily the Sinai-Ruelle-Bowen measure~\cite{You} as the
finite time response to an external perturbation does not require it;
an SRB measure is, however, necessary for the infinite time response
to be differentiable with respect to a deterministic external
perturbation~\cite{Rue1,Rue2}.

The difference between the new time-dependent ensemble
average of $A$ and its previous stationary (for the unperturbed
dynamics) value is then called the ``response'':
\begin{equation}
\label{eq:average_response}
\mbox{Response of }A=\delta\langle A\rangle(t)=\langle
A\rangle_{new}(t)-\langle A\rangle_{old}.
\end{equation}
Observe that above the average is taken not only with respect to the
statistical ensemble of solutions, but also over all possible
realizations of the external stochastic perturbation. Our goal here is
to derive the leading order term in the response, which depends only
on statistics of the unperturbed dynamical system, under the
assumption that the perturbation is sufficiently small.

The more obvious, ``brute force'' approach, would be to do the
following:
\begin{enumerate}[1.]
\item Start with a point $\BV x$ on $\MM$, and emit two trajectories
  out of $\BV x$: the unperturbed one, given by $\phi_t\BV x$, and the
  perturbed one, given by the corresponding solution of the perturbed
  system.
\item Clearly, both trajectories, perturbed and unperturbed, are
  generally nonlinear functions of elapsed time $t$ and initial
  condition $\BV x$. So, assuming that the unperturbed solution
  $\phi_t\BV x$ is ``known'', figure out a suitable way to ``expand''
  the perturbed solution around the unperturbed one in small
  increments, and keep only the leading order term.
\item Recall that this has to be done for every $\BV x\in\MM$, so,
  average the above result out with respect to the invariant measure
  $\mu$, and over all possible realizations of the external stochastic
  perturbation. Provided that the leading order response from the
  previous item is somehow expressed in terms of trajectories of the
  unperturbed system, the $\mu$-average can be replaced with the
  long-term time average, with help of Birkhoff's theorem~\cite{Bir}.
\end{enumerate}
With the exception of averaging across realizations of stochastic
perturbations, this is what was previously done for the deterministic
perturbations of chaotic and stochastic dynamical
systems~\cite{AbrMaj4,AbrMaj5,AbrMaj6,Abr5,Abr6,Abr7,Abr12}. It is a
long and cumbersome way of deriving the response, and the result
involves the (computationally expensive) tangent map $\BM T_{\BV
  x}^t$, given by
\begin{equation}
\label{eq:tangent_map}
\BM T_{\BV x}^t=\parderiv{}{\BV x}\phi_t\BV x.
\end{equation}
The situation is further complicated by the fact that, for chaotic
dynamical systems, $\BM T_{\BV x}^t$ grows exponentially fast in $t$,
which causes a numerical instability for moderate and long response
times. Thus, this approach can only be practically used for rather
short response times (although it is usually quite precise, provided
that the response time is sufficiently
short~\cite{Abr5,Abr6,Abr7,Abr12}).

\subsection{The forward Kolmogorov equation}

Another way to compute the average response is to employ the concept
of the probability density $p$ of a statistical ensemble distribution.
The key idea here is to use that fact that, while $\BV x_t$ is
governed by nonlinear dynamics, the partial differential equation for
$p$ (called the forward Kolmogorov equation~\cite{GikSko}, and also
the Fokker-Planck equation~\cite{Ris}) is linear. In particular, for
the deterministic dynamical system in~\eqref{eq:dyn_sys}, the forward
Kolmogorov equation for $p$ is given by
\begin{equation}
\label{eq:dyn_sys_kolm}
\parderiv{}t p(t,\BV x)=-D\cdot(p(t,\BV x)\BV f(\BV x)),
\end{equation}
where $D$ is the differentiation operator with respect to the
vector-argument of the function it acts upon. Above, the dot-product
of $D$ with a vector-function $\BV a(\BV x)$ refers to
\begin{equation}
D\cdot\BV a(\BV x)=\parderiv{a_i(\BV x)}{x_i},
\end{equation}
with the usual summation convention. Observe that in order for the
solution $p$ of~\eqref{eq:dyn_sys_kolm} to remain a probability
density, its integral over $\RN$ must remain equal to 1 (which,
together with the non-negativity of $p$, implies that $p$ must vanish
at infinity), even if $p$ itself changes with time. Indeed, one can
verify that the integral of $p$ over $\RN$ is preserved
by~\eqref{eq:dyn_sys_kolm}, which is necessary for $p$ to remain a
probability density.

The Kolmogorov equation above is an extremely useful tool for working
with the statistical properties of the system in \eqref{eq:dyn_sys},
since it describes the statistical distribution of the system in a
direct fashion. Unfortunately, it cannot be used directly to compute
the response of the deterministic system in~\eqref{eq:dyn_sys}, for
the following reason.

Since, as stated earlier, any solution of~\eqref{eq:dyn_sys} attracts
to $\MM$ as $t$ becomes infinite, it would be natural to think that,
in the limit as $t\to\infty$, $p$ becomes the density of the ergodic
invariant measure $\mu$ on $\MM$. However, here lies the fundamental
``incompatibility'' of the Kolmogorov equation
in~\eqref{eq:dyn_sys_kolm}, and the limiting dynamics
of~\eqref{eq:dyn_sys} on its global attractor $\MM$: for many applied
dynamical systems, especially those with dissipation and forcing
\cite{EckRue,Rue1,Rue2,You}, the invariant measure $\mu$ on $\MM$ is
not differentiable in $\BV x$ (it is also said that it is not
continuous with respect to the Lebesgue measure on $\MM$).  In this
situation, the (non-stationary) solution $p(t,\BV x)$
of~\eqref{eq:dyn_sys_kolm} contracts exponentially rapidly along
certain directions of the phase space (while appropriately expanding
transversally, so that its integral over $\RN$ remains 1), becoming
singular in the infinite time limit.

Observe that above we considered arguably the most simple setup for a
dynamical system, which describes a wide class of applied
problems. Yet, we cannot make use of the Kolmogorov
equation~\eqref{eq:dyn_sys_kolm} to statistically describe dynamics
near the attractor of the system in~\eqref{eq:dyn_sys}, which is
necessary for understanding of how the system responds to an external
perturbation. Therefore, in order to be able to use the Kolmogorov
equation in~\eqref{eq:dyn_sys_kolm}, we must be willing to consider a
suitable modification of~\eqref{eq:dyn_sys}, which renders its
invariant measure $\mu$ continuous with respect to the Lebesgue
measure on $\RN$. Arguably, the simplest such modification is achieved
via a stochastic noise added into the otherwise deterministic
dynamical system in~\eqref{eq:dyn_sys}.

\subsection{Stochastic dynamics}

Here we are going to consider a stochastic modification
of~\eqref{eq:dyn_sys}, achieved via introducing an additional noise
term via a Wiener process $\BV W_t$ of dimension $K$:
\begin{equation}
\label{eq:dyn_sys_Ito}
\dif\BV x_t=\BV f(\BV x_t)\dif t+\BM G(\BV x_t)\dif\BV W_t,
\end{equation}
where $\BM G(\BV x_t)$ is a smooth $N\times K$ matrix. For
convenience, here we interpret the resulting integral of the solution
\begin{equation}
\label{eq:Ito_integral}
\BV x_t=\BV x+\int_0^t\BV f(\BV x_s)\dif s+\int_0^t\BM G(\BV x_s)
\dif\BV W_s
\end{equation}
in the sense of It\^o~\cite{Ito,Ito2}. The forward Kolmogorov
equation for the differential equation in~\eqref{eq:dyn_sys_Ito} is
given by~\cite{GikSko,Oks}
\begin{equation}
\label{eq:dyn_sys_Ito_kolm}
\parderiv pt=-D\cdot(p\BV f)+\frac 12 D^2:(p\BM G\BM G^T),
\end{equation}
where ``$:$'' denotes the Frobenius product of two matrices, so that
\begin{equation}
D^2:\BM A(\BV x)=\parderiv{A_{ij}(\BV x)}{x_i\partial
  x_j}\quad\mbox{for an }N\times N\mbox{ matrix }\BM A.
\end{equation}
To ensure the smoothness of solutions of~\eqref{eq:dyn_sys_Ito_kolm},
here we follow~\cite{Pav} and assume that both $\BV f$ and $\BM G$
have bounded derivatives of all orders, and that the matrix product
$\BM G\BM G^T$ is uniformly positive definite in $\RN$. The latter
automatically means that the columns of $\BM G$ span $\RN$ for any
$\BV x\in\RN$, implying $K\geq N$. Additionally, we will assume that
there is a unique smooth stationary probability density $p_0$ which
sets the right-hand side of~\eqref{eq:dyn_sys_Ito_kolm} to zero.

Observe that the solution $\BV x_t$ of~\eqref{eq:dyn_sys_Ito} cannot
be represented by a $t$-semigroup like in~\eqref{eq:dyn_sys_soln},
since $\BV W_t$ depends on $t$ explicitly.  However, instead a similar
representation can be done for the Kolmogorov equation
in~\eqref{eq:dyn_sys_Ito_kolm} with help of the transitional
probability density $p^*$. Indeed, let $p^*(t,\BV x,\BV x_0)$ denote
the solution of~\eqref{eq:dyn_sys_Ito_kolm}, for which the initial
condition at $t=0$ is the delta-function $\delta(\BV x-\BV
x_0)$. Then, assuming that at time $t$ the solution is $p(t,\BV x)$,
its extension $p(t+s,\BV x)$ for $s\geq 0$ can be expressed as a
convolution with $p^*$ as follows:
\begin{equation}
\label{eq:kolm_semigroup}
p(t+s,\BV x)=\int_\RN p^*(s,\BV x,\BV y)p(t,\BV y)\dif\BV
y\eqdef\Lkolmsoln^s p(t,\BV x).
\end{equation}
Now, let $p_0$ denote the stationary smooth probability density
of~\eqref{eq:dyn_sys_Ito_kolm}, such that
\begin{equation}
\frac 12 D^2:(p_0\BM G\BM G^T)-D\cdot(p_0\BV f)=0,
\end{equation}
and, therefore,
\begin{equation}
p_0=\Lkolmsoln^t p_0\quad\mbox{ for any }t\geq 0.
\end{equation}
Then, the statistically average value of a function $A(\BV x)$ is
given by
\begin{equation}
\langle A\rangle=\int_\RN A(\BV x)p_0(\BV x)\dif\BV x,
\end{equation}
where we assume that $A(\BV x)$ is such that the integral above is
finite.  As before, if a statistical ensemble of solutions $\BV x_t$
of~\eqref{eq:dyn_sys_Ito} is distributed according to $p_0$, then the
ensemble average of $A$ is constant in time, even though each
individual solution in such an ensemble is in general not stationary.

\subsection{The layout of the paper}

In what is to follow, we arrange the presentation in the reverse order
(relative to what was presented above), due to the fact that, as was
mentioned previously, it turns out to be easier to start with a
stochastic differential equation of the form in~\eqref{eq:dyn_sys_Ito}
and derive the leading order response via the Kolmogorov
equation~\eqref{eq:dyn_sys_Ito_kolm}, which we do in
Section~\ref{sec:stochastic_dynamics}. Then, in
Section~\ref{sec:deterministic_dynamics} we return to the
deterministic unperturbed dynamics, and derive the response formula in
the ``brute force'' fashion, sketched above. In
Section~\ref{sec:equivalence} we show that, if one formally replaces
the invariant measure $\mu$ of the deterministic system
in~\eqref{eq:dyn_sys} with a smooth density approximation, then the
response formulas for the deterministic and stochastic unperturbed
dynamics become equivalent. In Section~\ref{sec:quasi-Gaussian} we
derive simplified response formulas for both the deterministic and
stochastic unperturbed dynamics under the assumption that the
probability density of the unperturbed state is Gaussian, as was
previously done in~\cite{AbrMaj4,AbrMaj5,AbrMaj6,MajAbrGro,AbrKje} for
deterministic perturbations. In
Section~\ref{sec:computational_experiments} we present the numerical
experiments with both the deterministic and stochastically forced
Lorenz 96 models to verify the computed response
formulas. Section~\ref{sec:summary} summarizes the results of the
work.

\section{Leading order response of a stochastic dynamics to a stochastic
perturbation}
\label{sec:stochastic_dynamics}

We start by considering the response of the stochastic dynamics
in~\eqref{eq:dyn_sys_Ito}, as due to the fact that the invariant state
$p_0$ of the unperturbed dynamics in~\eqref{eq:dyn_sys_Ito} is a
smooth stationary solution of the Kolmogorov equation
in~\eqref{eq:dyn_sys_Ito_kolm}, it is in fact much easier technically
to derive the corresponding leading order response formula (as opposed
to the situation with a deterministic unperturbed dynamics).  Here we
will consider two different types of perturbation: first, when the
existing stochastic matrix is perturbed, and, second, when a new,
statistically independent stochastic perturbation is added.

\subsection{Perturbing the existing stochastic term}

First, we are going to assume that the already present
in~\eqref{eq:dyn_sys_Ito} stochastic diffusion matrix $\BM G(\BV x)$
is perturbed by a small time-dependent term starting at $t=0$:
\begin{equation}
\dif\BV x_t=\BV f(\BV x_t)\dif t+(\BM G(\BV x_t)+\varepsilon\eta(t)
\BM H(\BV x_t))\dif\BV W_t,
\end{equation}
where $0<\varepsilon\ll 1$ is a small constant parameter to signify
that the perturbation is small, $\BM H(\BV x)$ is a matrix of the same
dimension and smoothness properties as $\BM G$, and $\eta(t)$ is a
bounded, piecewise continuous and square-integrable function, which is
zero for negative values of $t$. Then, the corresponding perturbed
Kolmogorov equation is obtained from~\eqref{eq:dyn_sys_Ito_kolm} by
replacing $\BM G$ with $\BM G+\varepsilon\eta\BM H$:
\begin{equation}
\parderiv {p^\varepsilon}t = -D\cdot(p^\varepsilon\BV f)+\frac 12 D^2
:\left(p^\varepsilon(\BM G+\varepsilon\eta(t)\BM H)(\BM G+\varepsilon
\eta(t)\BM H)^T\right).
\end{equation}
We assume that the solution $p^\varepsilon$ of the perturbed
Kolmogorov equation above depends smoothly on $\varepsilon$ for
sufficiently small $\varepsilon$, and admits the expansion
\begin{equation}
p^\varepsilon=p_0+\varepsilon p_1+\varepsilon^2 p_2+\ldots,
\end{equation}
where $p_0$ is the stationary solution of the unperturbed Kolmogorov
equation, while the expansion terms $p_i$, $i>0$ are independent on
$\varepsilon$ and have zero initial conditions. We seek the perturbed
solution in the leading order of $\varepsilon$, which leads directly
to
\begin{equation}
\parderiv {p_1}t=-D\cdot(p_1\BV f)+\frac 12 D^2:(p_1\BM G\BM G^T)
+\frac{\eta(t)}2 D^2:\left(p_0(\BM G\BM H^T+\BM H\BM G^T)\right).
\end{equation}
One can verify directly that the solution for $p_1$ is given by
\begin{equation}
p_1(t)=\frac 12\int_0^t\Lkolmsoln^{t-s}\left(D^2:(p_0(\BM G\BM H^T+\BM
H\BM G^T))\right)\eta(s)\dif s,
\end{equation}
where $\Lkolmsoln^t$ is defined in~\eqref{eq:kolm_semigroup}. The
response of $A$ in the leading order of $\varepsilon$ is thus given by
\begin{equation}
\delta\langle A\rangle(t)=\int_\RN\!\! A(p-p_0)\dif x =
\varepsilon\int_\RN\!\! Ap_1\dif x+O(\varepsilon^2)=\varepsilon
\int_0^t R_1(t-s)\eta(s)\dif s+O(\varepsilon^2),
\end{equation}
with
\begin{equation}
R_1(t)=\frac 12\int_\RN A(\BV x)\Lkolmsoln^t\left(D^2:(p_0(\BM
G\BM H^T+\BM H\BM G^T))\right)(\BV x)\dif\BV x.
\end{equation}
At this point, we use the definition of $\Lkolmsoln^t$
in~\eqref{eq:kolm_semigroup} to obtain $R_1(t)$ in terms of the
time-correlation function~\cite{Ris,Pav}
\begin{equation}
\label{eq:resp_corr_func}
R_1(t)=\frac 12\int_\RN\int_\RN A(\BV x)p^*(t,\BV x,\BV y)B(\BV
y)p_0(\BV y)\dif\BV y\dif\BV x,
\end{equation}
with
\begin{equation}
B(\BV y)=p_0^{-1}(\BV y)D^2:(p_0(\BM G\BM H^T+\BM H\BM G^T))(\BV y).
\end{equation}
Above, the division by $p_0$ is allowed since it is the solution to an
elliptic equation which vanishes at infinity, and thus is never zero
for finite $\BV y$.

For the practical computation of the time correlation function
in~\eqref{eq:resp_corr_func}, we use the Birkhoff-Khinchin
theorem~\cite{GikSko1} and replace the spatial integrals
in~\eqref{eq:resp_corr_func} with the time average along a long-term
trajectory as
\begin{equation}
R_1(t)=\lim_{r\to\infty}\frac 1{2r}\int_0^rA(\BV x_{t+s})B(\BV
x_s)\dif s,
\end{equation}
which results, after substituting the expression for $B(\BV x_s)$, in
\begin{equation}
\label{eq:R_1}
R_1(t)=\lim_{r\to\infty}\frac 1{2r}\int_0^rA(\BV x_{t+s})p_0^{-1}(\BV
x_s)D^2:\left(p_0(\BV x_s)(\BM G(\BV x_s)\BM H^T(\BV x_s)+\BM H(\BV
x_s)\BM G^T(\BV x_s))\right)\dif s.
\end{equation}
The long-term trajectory $\BV x_s$ above is computed via a numerical
simulation of~\eqref{eq:dyn_sys_Ito}, from an arbitrary initial
condition.

\subsection{Adding a new stochastic term}

Here we assume that a new small stochastic term is added
to~\eqref{eq:dyn_sys_Ito},
\begin{equation}
\dif\BV x_t=\BV f(\BV x_t)\dif t+\BM G(\BV x_t)\dif\BV W_t+\varepsilon
\eta(t)\BM H(\BV x_t)\dif\BV W_t',
\end{equation}
where the Wiener process $\BV W_t'$ is independent of $\BV W_t$. In
order to derive the corresponding Kolmogorov equation, we rewrite the
above equation in the form
\begin{equation}
\dif\BV x_t=\BV f(\BV x_t)\dif t+\widetilde{\BM G}(t,\BV x_t) \dif
\widetilde{\BV W}_t,
\end{equation}
where
\begin{equation}
\widetilde{\BV W}_t=\left(\begin{array}{c}\BV W_t \\ \BV W_t'
\end{array}\right),\qquad\widetilde{\BM G}(t,\BV x_t)=\Big(
\BM G(\BV x_t) \Big| \varepsilon \eta(t)\BM H(\BV x_t)\Big).
\end{equation}
The corresponding perturbed Kolmogorov equation is, obviously, given
by
\begin{equation}
\parderiv{p^\varepsilon}t = -D\cdot(p^\varepsilon\BV f)+\frac 12
D^2:(p^\varepsilon\widetilde{\BM G}\widetilde{\BM G}^T).
\end{equation}
Further observing that
\begin{equation}
\widetilde{\BM G}\widetilde{\BM G}^T=\BM G\BM G^T+\varepsilon^2
\eta^2(t)\BM H\BM H^T,
\end{equation}
we arrive at
\begin{equation}
\parderiv{p^\varepsilon}t = -D\cdot(p^\varepsilon\BV f)+\frac 12
D^2:(p^\varepsilon\BM G\BM G^T)+\frac{\varepsilon^2\eta^2(t)}2
D^2:(p^\varepsilon\BM H\BM H^T).
\end{equation}
Observe that there is no first-order term in $\varepsilon$, so we can
expand $p^\varepsilon$ near $p_0$ in even powers of $\varepsilon$ as
\begin{equation}
p=p_0+\varepsilon^2p_2+\varepsilon^4p_4+\ldots,
\end{equation}
where, as before, $p_{2i}$ for $i>0$ are independent of $\varepsilon$
and have zero initial condition, and similarly obtain the equation for
$p_2$ as
\begin{equation}
\parderiv {p_2}t=-D\cdot(p_2\BV f)+\frac 12 D^2:(p_2\BM G\BM G^T)+
\frac{\eta^2(t)}2 D^2:(p_0\BM H\BM H^T).
\end{equation}
Again, one can verify that $p_2$ is given by
\begin{equation}
p_2(t)=\frac 12\int_0^t\Lkolmsoln^{t-s} D^2:(p_0\BM H\BM H^T)
\eta^2(s)\dif s.
\end{equation}
The response of $A$ in the leading order of $\varepsilon$ (which is
now $\varepsilon^2$) is thus given by
\begin{equation}
\delta\langle A\rangle(t)=\int_\RN\!\! A(p-p_0)\dif x=\varepsilon^2
\int_\RN\!\! Ap_2\dif x+O(\varepsilon^4)=\varepsilon^2\int_0^t R_2(t-s)
\eta^2(s)\dif s+O(\varepsilon^4),
\end{equation}
with
\begin{equation}
R_2(t)=\frac 12\int_\RN A(\BV x)\Lkolmsoln^t\left(D^2:(p_0\BM H\BM
H^T)\right)(\BV x)\dif\BV x.
\end{equation}
Following the same steps as above for $R_1(t)$, we express $R_2(t)$
via the time average as
\begin{equation}
\label{eq:R_2}
R_2(t)=\lim_{r\to\infty}\frac 1{2r} \int_0^r A(\BV x_{t+s}) p_0^{-1}
(\BV x_s)D^2:\left(p_0(\BV x_s)\BM H(\BV x_s)\BM H^T(\BV x_s)
\right)\dif s.
\end{equation}
Observe that in this case the response is quadratic in
$\varepsilon\eta(t)$ (which is unlike the previous case, where the
existing diffusion matrix was perturbed).

\section{Leading order response of a deterministic dynamics to a
stochastic perturbation}
\label{sec:deterministic_dynamics}

Now, we consider a small external stochastic perturbation of the
deterministic system in~\eqref{eq:dyn_sys} of the form
\begin{equation}
\label{eq:dyn_sys_pert}
\dif\BV x_t=\BV f(\BV x_t)\dif t+\varepsilon\eta(t)\BM H(\BV x_t)\dif\BV
W_t,
\end{equation}
where the perturbation term has the same properties as in the previous
section, while, for the purposes of the derivation, we additionally
require $\BV f$ to be uniformly Lipschitz in $\RN$~\cite{GikSko,Pav}
to ensure the existence of solutions to~\eqref{eq:dyn_sys_pert}
(recall that the unperturbed system back in~\eqref{eq:dyn_sys} does
not necessarily require it for global existence).  Here, however, we
cannot use an expansion of Kolmogorov equation near the stationary
unperturbed state, since this state may not necessarily be continuous
with respect to the Lebesgue measure. Instead, we will have to employ
the differentiability of the resulting stochastic flow with respect to
the perturbation~\cite{Kun}.

We denote the solution to the perturbed system
in~\eqref{eq:dyn_sys_pert} by $\BV x_t^\varepsilon
=\phi_t^\varepsilon\BV x$, and rewrite \eqref{eq:dyn_sys_pert} in the
integral form as
\begin{equation}
\phi_t^\varepsilon\BV x=\BV x+\int_0^t\BV f(\phi_s^\varepsilon\BV x)
\dif s+\varepsilon\int_0^t\eta(s)\BM H(\phi_s^\varepsilon\BV x)\dif\BV
W_s,
\end{equation}
where the stochastic integral is computed in the sense of It\^o. Let
$A(\BV x)$ be a twice differentiable function, then one can write its
second-order Taylor expansion in $\varepsilon$ as
\begin{multline}
\label{eq:A_Taylor}
A(\phi_t^\varepsilon\BV x)-A(\phi_t\BV x)=\varepsilon DA(\phi_t\BV
x)\cdot\partial_\varepsilon\phi_t\BV x+\\+\frac{\varepsilon^2}2
\left[DA(\phi_t\BV x)\cdot\partial_\varepsilon^2\phi_t\BV
  x+D^2A(\phi_t\BV x):\left(\partial_\varepsilon\phi_t\BV
  x\otimes\partial_\varepsilon\phi_t\BV
  x\right)\right]+o(\varepsilon^2),
\end{multline}
where ``$\otimes$'' is the outer product of two vectors, that is,
\begin{equation}
\BV x\otimes\BV y=x_i y_j.
\end{equation}
Also, the following notation is used above:
\begin{equation}
\partial_\varepsilon\phi_t\BV x\eqdef\left.\parderiv{
  \phi_t^\varepsilon\BV x}\varepsilon\right|_{\varepsilon=0}.
\end{equation}
For the $\varepsilon$-derivative of $\phi_t^\varepsilon\BV x$ (which
we assume to exist almost surely for finite~$t$ according
to~\cite{Kun}) we compute
\begin{multline}
\label{eq:deriveps}
\parderiv{}\varepsilon\phi_t^\varepsilon\BV x=\int_0^tD\BV
f(\phi_s^\varepsilon\BV x)\parderiv{}\varepsilon\phi_s^\varepsilon\BV
x\dif s+\varepsilon\int_0^t\eta(s)D\BM H(\phi_s^\varepsilon\BV x)
\parderiv{}\varepsilon\phi_s^\varepsilon\BV x\,\dif\BV W_s+\\+\int_0^t
\eta(s)\BM H(\phi_s^\varepsilon\BV x)\dif\BV W_s,
\end{multline}
which results, by setting $\varepsilon=0$, in
\begin{equation}
\label{eq:varphix}
\partial_\varepsilon\phi_t\BV x=\int_0^tD\BV f(\phi_s\BV
x)\partial_\varepsilon\phi_s\BV x\dif s+\int_0^t\eta(s)\BM H(\phi_s\BV
x)\dif\BV W_s.
\end{equation}
At this point, we need to solve the It\^o integral equation above.
Applying the It\^o differentiation formula to both sides
of~\eqref{eq:varphix} results in
\begin{equation}
\dif(\partial_\varepsilon\phi_t\BV x)=D\BV f(\phi_t\BV x)
\partial_\varepsilon\phi_t\BV x\dif t+\eta(t)\BM H(\phi_t\BV x)\dif\BV
W_t.
\end{equation}
At the same time, it is easy to verify that the tangent map $\BM
T_{\BV x}^t$ from~\eqref{eq:tangent_map} satisfies
\begin{equation}
\label{eq:tmap}
\parderiv{}t\BM T_{\BV x}^t=D\BV f(\phi_t\BV x)\BM T_{\BV x}^t,
\end{equation}
which further yields
\begin{equation}
\dif(\partial_\varepsilon\phi_t\BV x)=\left(\parderiv{}t\BM T_{\BV
  x}^t\right)\left(\BM T_{\BV x}^t\right)^{-1}\partial_\varepsilon
\phi_t\BV x\dif t+\eta(t)\BM H(\phi_t\BV x)\dif\BV W_t.
\end{equation}
Now we multiply both sides of the above identity by the inverse of
$\BM T_{\BV x}^t$ on the left, which results, after taking into
account the identity
\[
\BM 0=\BM A^{-1}\parderiv{\BM I}t=\BM A^{-1}\parderiv{}t(\BM A\BM
A^{-1})=\BM A^{-1}\parderiv{ \BM A}t\BM A^{-1}+\parderiv{}t(\BM
A^{-1})
\]
for an arbitrary matrix $\BM A$, in
\begin{equation}
\left(\BM T_{\BV x}^t\right)^{-1}\dif(\partial_\varepsilon\phi_t\BV x)
=-\left(\parderiv{}t\left(\BM T_{\BV x}^t\right)^{-1}\right)\partial_\varepsilon
\phi_t\BV x\dif t+\eta(t)\left(\BM T_{\BV x}^t\right)^{-1}\BM
H(\phi_t\BV x)\dif\BV W_t.
\end{equation}
Pulling the first term in the right-hand side above to the left,
combining the terms and integrating from 0 to $t$, we
arrive at
\begin{equation}
\label{eq:Tphi}
\left(\BM T_{\BV x}^t\right)^{-1}\partial_\varepsilon\phi_t\BV x=
\int_0^t\eta(s)\left(\BM T_{\BV x}^s\right)^{-1}\BM H(\phi_s\BV x)\dif\BV
W_s+\left[\left(\BM T_{\BV x}^t\right)^{-1},
  \partial_\varepsilon\phi_t\BV x\right]_0^t,
\end{equation}
where the last term is the quadratic covariation of the processes
$\left(\BM T_{\BV x}^t\right)^{-1}$ and $\partial_\varepsilon\phi_t\BV
x$:
\begin{equation}
\label{eq:quad_cov}
\left[\left(\BM T_{\BV x}^t\right)^{-1},\partial_\varepsilon\phi_t\BV
  x\right]_0^t=\int_0^t\dif\left(\left(\BM T_{\BV x}^s\right)^{-1}
\right)\dif\left(\partial_\varepsilon\phi_s\BV x\right)=\int_0^t
\parderiv{}s\left(\left(\BM T_{\BV x}^s\right)^{-1}\right)\dif s\;
\dif\left(\partial_\varepsilon\phi_s\BV x\right).
\end{equation}
However, since we have assumed above that
$\partial_\varepsilon\phi_t\BV x= \int_0^t\dif \left(
\partial_\varepsilon\phi_s\BV x\right)$ is almost surely finite for
finite $t$, the quadratic covariation above in~\eqref{eq:quad_cov} is
almost surely zero. Further multiplying~\eqref{eq:Tphi} by $\BM
T_{\BV x}^t$ on the left and taking into account its cocycle property,
we finally arrive at
\begin{equation}
\partial_\varepsilon\phi_t\BV x=\int_0^t\eta(s)\BM T_{\phi_s\BV
  x}^{t-s} \BM H(\phi_s\BV x)\dif\BV W_s.
\end{equation}
For the second $\varepsilon$-derivative we further obtain by
the differentiation of~\eqref{eq:deriveps}
\begin{multline}
\parderiv{^2}{\varepsilon^2}\phi_t^\varepsilon\BV x=\int_0^t\left[D\BV
  f(\phi_s^\varepsilon\BV x)\parderiv{^2}{\varepsilon^2}
  \phi_s^\varepsilon\BV x+D^2\BV f(\phi_s^\varepsilon\BV x)
  :\left(\parderiv{}\varepsilon\phi_s^\varepsilon\BV x\otimes
  \parderiv{}\varepsilon\phi_s^\varepsilon\BV x\right)\right]\dif s
+\\+\int_0^t\left[2\eta(s)D\BM H(\phi_s^\varepsilon\BV x)\parderiv{}
  \varepsilon\phi_s^\varepsilon\BV x+\varepsilon\eta(s)\parderiv{}
  \varepsilon\left(D\BM H(\phi_s^\varepsilon\BV x)\parderiv{}
  \varepsilon\phi_s^\varepsilon\BV x\right)\right]\dif\BV W_s,
\end{multline}
which becomes, upon setting $\varepsilon=0$,
\begin{multline}
\partial_\varepsilon^2\phi_t\BV x=\int_0^t\left[D\BV f(\phi_s\BV x)
  \partial_\varepsilon^2\phi_s\BV x+D^2\BV f(\phi_s\BV x):\left(
  \partial_\varepsilon\phi_s\BV x\otimes\partial_\varepsilon\phi_s\BV
  x\right)\right]\dif s +\\+2\int_0^t\eta(s)D\BM H(\phi_s\BV x)
\partial_\varepsilon\phi_s\BV x\dif\BV W_s.
\end{multline}
Now recall that, according to the definition of the average response
in~\eqref{eq:average_response}, we should compute the expectation
(that is, the average) of the second-order Taylor expansion
in~\eqref{eq:A_Taylor} over all realizations of $\BV W_t$, which leads
to
\begin{multline}
\expec A(\phi_t^\varepsilon\BV x)-A(\phi_t\BV x)=\varepsilon DA(\phi_t
\BV x)\cdot\expec\partial_\varepsilon\phi_t\BV x+\\+\frac{
  \varepsilon^2}2\left[DA(\phi_t\BV x)\cdot\expec
  \partial_\varepsilon^2\phi_t\BV x+D^2A(\phi_t\BV x):\expec \left(
  \partial_\varepsilon\phi_t\BV x\otimes\partial_\varepsilon\phi_t\BV
  x\right)\right]+o(\varepsilon^2).
\end{multline}
One immediately observes that
\begin{subequations}
\begin{equation}
\expec\partial_\varepsilon\phi_t\BV x=\expec\int_0^t\eta(s)\BM
T_{\phi_s\BV x}^{t-s}\BM H(\phi_s\BV x)\dif\BV W_s=0,
\end{equation}
\begin{equation}
\expec\int_0^t\eta(s)D\BM H(\phi_s\BV x)\partial_\varepsilon\phi_s\BV
x\dif\BV W_s=0,
\end{equation}
\end{subequations}
where in the first identity the integrand is not a random variable,
while in the second the integrand is adapted to the natural filtration
of $\BV W_t$. After some computation while remembering Duhamel's
principle and It\^o's isometry, we arrive at
\begin{subequations}
\begin{equation}
\expec\left(\partial_\varepsilon\phi_t\BV
x\otimes\partial_\varepsilon\phi_t\BV x\right)=\int_0^t\BM T_{\phi_s
  \BV x}^{t-s}\BM H(\phi_s\BV x)\BM H^T(\phi_s\BV x)\left(\BM
T_{\phi_s\BV x}^{t-s}\right)^T\eta^2(s)\dif s,
\end{equation}
\begin{equation}
\expec\partial_\varepsilon^2\phi_t\BV x=\int_0^t\int_0^s\BM
T_{\phi_s\BV x}^{t-s}D^2\BV f(\phi_s\BV x):\left[\BM T_{\phi_\tau\BV
    x}^{s-\tau}\BM H(\phi_\tau\BV x)\BM H^T(\phi_\tau\BV x)\left(\BM
  T_{\phi_\tau\BV x}^{s-\tau}\right)^T \right]\eta^2(\tau)\dif\tau\dif
s.
\end{equation}
\end{subequations}
Now we recall that the expectations above are taken under the
condition that the stochastic flows start at $\BV x$, which, in turn,
belongs to the attractor of~\eqref{eq:dyn_sys}. Therefore, we further
need to average the result above over the invariant measure $\mu$ of
the unperturbed system. The result is, after discarding the
higher-order terms,
\begin{subequations}
\begin{equation}
\delta\langle A\rangle(t)=\varepsilon^2\int_0^t R(t-s)\eta^2(s)\dif s+
o(\varepsilon^2),
\end{equation}
\begin{multline}
\label{eq:RR}
R(t)=\frac 12\int_\MM\bigg[D^2A(\phi_t\BV x):\left(\BM T_{\BV x}^t\BM
  H(\BV x)\BM H^T(\BV x)\left(\BM T_{\BV x}^t\right)^T\right)+\\+
  DA(\phi_t\BV x)\int_0^t\BM T_{\phi_s\BV x}^{t-s} D^2\BV f(\phi_s\BV
  x):\left(\BM T_{\BV x}^s\BM H(\BV x)\BM H^T(\BV x)\left(\BM T_{\BV
    x}^s\right)^T\right)\dif s\bigg]\dif\mu(\BV x).
\end{multline}
\end{subequations}

\section{The equivalence of the response formulas for the
deterministic and stochastic dynamics}
\label{sec:equivalence}

While the response formula in~\eqref{eq:RR} is rather complicated for
a practical use, one can actually show that the response operator
$R(t)$ in~\eqref{eq:RR} can be written in a more concise way:
\begin{equation}
\label{eq:R}
R(t)=\frac 12\int_\MM \parderiv{^2A(\phi_t\BV x)}{\BV x^2}:\big(\BM
H(\BV x)\BM H^T(\BV x)\big)\dif\mu(\BV x).
\end{equation}
Indeed, observe that, first,
\begin{subequations}
\begin{equation}
\parderiv{A(\phi_t\BV x)}{\BV x}=DA(\phi_t\BV x)\BM T_{\BV x}^t,
\end{equation}
\begin{equation}
\parderiv{^2A(\phi_t\BV x)}{\BV x^2}=D^2A(\phi_t\BV x):\left(\BM
T_{\BV x}^t\otimes\BM T_{\BV x}^t\right) +DA(\phi_t\BV x)\parderiv{}
{\BV x}\BM T_{\BV x}^t,
\end{equation}
\end{subequations}
where the combination of the Frobenius and outer product for matrices
denotes
\begin{equation}
\big[\BM A:(\BM B\otimes\BM C)\big]_{ij}=A_{kl}B_{ki}C_{lj}.
\end{equation}
Next, differentiating~\eqref{eq:tmap} with respect to $\BV x$ yields
\begin{equation}
\parderiv{}t\left(\parderiv{}{\BV x}\BM T_{\BV x}^t\right)=D\BV
f(\phi_t\BV x)\parderiv{}{\BV x}\BM T_{\BV x}^t+D^2\BV f(\phi_t\BV
x):\left(\BM T_{\BV x}^t\otimes\BM T_{\BV x}^t\right).
\end{equation}
Duhamel's principle then yields
\begin{equation}
\parderiv{}{\BV x}\BM T_{\BV x}^t=\int_0^t\BM T_{\phi_s\BV x}^{t-s}
D^2\BV f(\phi_s\BV x):\left(\BM T_{\BV x}^s\otimes\BM T_{\BV
  x}^s\right)\dif s.
\end{equation}
Combining the results, we obtain
\begin{equation}
\parderiv{^2A(\phi_t\BV x)}{\BV x^2}=D^2A(\phi_t\BV x): \left(\BM
T_{\BV x}^t\otimes\BM T_{\BV x}^t\right) +DA(\phi_t\BV x)\int_0^t\BM
T_{\phi_s\BV x}^{t-s} D^2\BV f(\phi_s\BV x):\left(\BM T_{\BV
  x}^s\otimes\BM T_{\BV x}^s\right)\dif s,
\end{equation}
which leads to the above claim.

It is interesting that a different response formula was obtained
in~\cite{Luc} for a stochastic perturbation of a deterministic
dynamics.\footnote{The author privately communicated that his response
  formula applies to a perturbation noise in the Stratonovich form.}
The derivation in~\cite{Luc} was also different: instead, a stochastic
perturbation was used directly in the second-order response formula
for deterministic perturbations~\cite{Rue4}, which was further scaled
by a factor of one-half.

As we mentioned above, generally one cannot assume that the invariant
measure of the deterministic dynamics of the form~\eqref{eq:dyn_sys}
possesses a density, since most often the compact set on which the
solution of~\eqref{eq:dyn_sys} lives has a complicated structure.
However, let us assume that there exists a smooth probability density
$p_0(\BV x)>0$, such that $p_0\dif\BV x$ is a suitable, in an
appropriate for our purposes sense, approximation for the invariant
measure $\dif\mu$. Under such a hypothetical assumption, one
writes~\eqref{eq:R} in the form
\begin{equation}
R(t)=\frac 12\int_\MM \parderiv{^2A(\phi_t\BV x)}{\BV x^2}:\big(\BM
H(\BV x)\BM H^T(\BV x)\big)p_0(\BV x)\dif\BV x.
\end{equation}
Now that $\mu$ is replaced by the density $p_0(\BV x)$, one can
integrate the above expression by parts, obtaining
\begin{equation}
R(t)=\frac 12\int_\MM A(\phi_t\BV x) D^2:\left(p_0(\BV x)\BM H(\BV
x)\BM H^T(\BV x)\right)\dif\BV x.
\end{equation}
Replacing the measure averages with the time averages with help of
Birkhoff's theorem, we obtain the same formula as in~\eqref{eq:R_2}:
\begin{equation}
\label{eq:R_3}
R(t)=\lim_{r\to\infty}\frac 1{2r}\int_0^rA(\BV x_{t+s})p_0^{-1}(\BV
x_s)D^2:\left(p_0(\BV x_s)\BM H(\BV x_s)\BM H^T(\BV x_s)\right)\dif s.
\end{equation}
In other words, under the assumption of a differentiable approximation
to the invariant state, the time-averaged response formula for the
deterministic unperturbed dynamics is identical to the response
formula for the stochastic dynamics in~\eqref{eq:R_2}, where the
external stochastic perturbation is statistically independent to the
unperturbed noise term.

Below we will see that this approach allows to obtain a sensible
approximation to the response operator even in a situation where the
unperturbed dynamics is purely deterministic, similar to what was
observed for the deterministic perturbations
in~\cite{AbrMaj4,AbrMaj5,AbrMaj6,MajAbrGro}.

\section{The quasi-Gaussian approximation for the response operator}
\label{sec:quasi-Gaussian}

Observe that the response formulas for the stochastic unperturbed
dynamics in \eqref{eq:R_1} and~\eqref{eq:R_2} are not computable
directly, since the equilibrium density $p_0(\BV x)$
of~\eqref{eq:dyn_sys_Ito} is not generally known explicitly. It is,
theoretically, possible to compute the response in~\eqref{eq:R} by
computing the tangent map $\BM T_{\BV x}^t$ in parallel with $\BV x_t$
(for more details, see
\cite{AbrMaj4,AbrMaj5,AbrMaj6,Abr5,Abr6,Abr7,Abr12}) and
using~\eqref{eq:RR} expressed as a time-lagged autocorrelation
function over the time-series average. However, the latter option is
very expensive from the computational standpoint, and, for a chaotic
unperturbed dynamical system, it will only remain computationally
stable for short response times.

Instead, we are going to use a simplified method to compute the
response, called the quasi-Gaussian FDT (qG-FDT)
approximation~\cite{AbrMaj4}.  The main idea of the qG-FDT
approximation is that $p_0(\BV x)$ in~\eqref{eq:R_1}, \eqref{eq:R_2}
and~\eqref{eq:R_3} is replaced with its Gaussian approximation, which
has the same mean state and covariance matrix as does $p_0(\BV
x)$. For that, first observe that~\eqref{eq:R_1} and~\eqref{eq:R_2}
(which is identical to~\eqref{eq:R_3}) can be written as
\begin{subequations}
\begin{multline}
\label{eq:R_1_2}
R_1(t)=\lim_{r\to\infty}\frac 1{2r}\int_0^rA(\BV x_{t+s}) D^2:\left(
\BM G(\BV x_s)\BM H^T(\BV x_s)+\BM H(\BV x_s)\BM G^T(\BV x_s)\right)
\dif s+\\+\lim_{r\to\infty}\frac 1r\int_0^rA(\BV x_{t+s})\left(D\cdot
\left(\BM G(\BV x_s)\BM H^T(\BV x_s)+\BM H(\BV x_s)\BM G^T(\BV x_s)
\right)\right)\cdot \frac{D p_0(\BV x_s)}{p_0(\BV x_s)}\dif s+\\+
\lim_{r\to\infty}\frac 1{2r}\int_0^rA(\BV x_{t+s})\left(\BM G(\BV x_s)
\BM H^T(\BV x_s)+\BM H(\BV x_s)\BM G^T(\BV x_s)\right):\frac{D^2
  p_0(\BV x_s)}{p_0(\BV x_s) }\dif s.
\end{multline}
\begin{multline}
\label{eq:R_2_2}
R_2(t)=\lim_{r\to\infty}\frac 1{2r} \int_0^r A(\BV x_{t+s}) D^2:
\left(\BM H(\BV x_s)\BM H^T(\BV x_s)\right)\dif s+\\+
\lim_{r\to\infty}\frac 1 r\int_0^r A(\BV x_{t+s})\left(D\cdot\left(\BM
H(\BV x_s)\BM H^T(\BV x_s)\right)\right)\cdot\frac{D p_0(\BV x_s)}{
  p_0(\BV x_s)} \dif s+\\+\lim_{r\to\infty}\frac 1{2r}\int_0^r A(\BV
x_{t+s})\left( \BM H(\BV x_s)\BM H^T(\BV x_s)\right):\frac{D^2 p_0(\BV
  x_s)}{p_0(\BV x_s)}\dif s.
\end{multline}
\end{subequations}
Now, let us denote the mean state of $p_0(\BV x)$ as $\BV m$, and its
covariance matrix as $\BM C$:
\begin{subequations}
\begin{equation}
\BV m=\int_\RN \BV x\, p_0(\BV x)\dif\BV x=\lim_{r\to\infty}\frac
1r\int_0^r\BV x_s\dif\BV s,
\end{equation}
\begin{equation}
\BM C=\int_\RN(\BV x-\BV m)\otimes(\BV x-\BV m)\,p_0(\BV x)\dif\BV x=
\lim_{r\to\infty}\frac 1r\int_0^r(\BV x_s-\BV m)\otimes(\BV x_s-\BV
m)\dif\BV s.
\end{equation}
\end{subequations}
Then, the Gaussian approximation $p_0^G(\BV x)$ for $p_0(\BV x)$ is
given by the explicit formula
\begin{equation}
p_0^G(\BV x)=\frac 1{\sqrt{(2\pi)^N\det{\BM C}}}\exp\left(-\frac 12
(\BV x-\BV m)\cdot\BM C^{-1}(\BV x-\BV m)\right),
\end{equation}
which results in
\begin{subequations}
\label{eq:Gauss_app}
\begin{equation}
\frac{D p_0^G(\BV x)}{p_0^G(\BV x)}=-\BM C^{-1}(\BV x-\BV m),
\end{equation}
\begin{equation}
\frac{D^2 p_0^G(\BV x)}{p_0^G(\BV x)}=\BM C^{-1}\big((\BV x-\BV m)
\otimes(\BV x-\BV m)\big)\BM C^{-1}-\BM C^{-1}.
\end{equation}
\end{subequations}
The approximations above are then inserted directly
into~\eqref{eq:R_1_2} and~\eqref{eq:R_2_2}, resulting in explicit
time-lagged autocorrelation functions, computed along the long-term
time series of solutions of the unperturbed system
in~\eqref{eq:dyn_sys_Ito}. Observe that the autocorrelations
in~\eqref{eq:R_1_2} and~\eqref{eq:R_2_2}, computed via the Gaussian
approximations in~\eqref{eq:Gauss_app}, simplify somewhat further if
the matrices $\BM G$ and $\BM H$ are constant, as only one term out of
the three remains in such a case.

\section{Computational experiments}
\label{sec:computational_experiments}

In this section we investigate the validity of the qG-FDT response
formulas in~\eqref{eq:R_1_2} and~\eqref{eq:R_2_2} for both the
deterministic and stochastic unperturbed dynamics. For the test-bed
dynamical system, we choose the well-known Lorenz 96 model.

\subsection{The rescaled Lorenz 96 model}

The test system used to study the new method of stochastic
parameterization here is the rescaled Lorenz 96 system with stochastic
forcing. Without the stochastic forcing, it was previously used in
\cite{Abr8,Abr9,Abr10,Abr11,Abr12,AbrKje,LucSar} to study the
deterministic reduced model parameterization, as well as the average
response to deterministic external perturbations. The rescaled Lorenz
96 system with stochastic forcing is given by
\begin{equation}
\label{eq:lorenz_rescaled}
\dif x_i=\left[x_{i-1}(x_{i+1}-x_{i-2})+\frac 1\beta(\bar x(x_{i+1}-x_{i-2})
-x_i)+\frac{F-\bar x}{\beta^2}\right]\dif t+G\,\dif W_t^i,
\end{equation}
with $1\leq i\leq N$. Above, $W_t^i$ denotes the family of $N$
mutually independent Wiener processes, indexed by $i$, with $G$ being
a constant stochastic forcing parameter, so that, in terms of the
notations in~\eqref{eq:dyn_sys_Ito}, $\BM G(\BV x)$ is a constant
multiple of the identity matrix,
\begin{equation}
\BM G(\BV x)=G\BM I.
\end{equation}
The model has periodic boundary conditions: $x_{i+N}=x_i$. The
parameters $\bar x$ and $\beta$ are the statistical mean and the
standard deviation, respectively, for the corresponding unrescaled
Lorenz 96 model \cite{Lor,LorEma}
\begin{equation}
\deriv{x_i}t=x_{i-1}(x_{i+1}-x_{i-2})-x_i+F,
\end{equation}
with the same periodic boundary conditions. The rescaling above
ensures that, in the absence of the stochastic forcing (that is,
$G=0$) the Lorenz 96 model in \eqref{eq:lorenz_rescaled} has zero mean
state and unit standard deviation, and that the time scale of
evolution of its solution is roughly the same for different values of
$F$. In the current work, we test the response theory, developed
above, for two values of $F$, $F=24$ and $F=8$, and two values of $G$,
$G=0.5$ and $G=0$, with the latter corresponding to the purely
deterministic unperturbed dynamics. As in the original
paper~\cite{LorEma}, we set $N=40$.

We must note that the right-hand side of~\eqref{eq:lorenz_rescaled}
does not satisfy the requirements imposed on~\eqref{eq:dyn_sys_Ito} in
Section~\ref{sec:intro}; observe that the deterministic part of the
right-hand side of~\eqref{eq:lorenz_rescaled} is neither bounded nor
even uniformly Lipschitz in~$\RN$, which means that the existence of
strong solutions to~\eqref{eq:lorenz_rescaled} is not
guaranteed~\cite{GikSko,Pav}. Nonetheless, below we demonstrate via
the numerical simulations that, for the chosen parameters of the
system, numerical solutions exist for a long enough time to allow the
reliable time-averaging for the response computation.

\subsection{Long-term statistics of the unperturbed dynamics}
\label{sec:stats}

\begin{figure}%
\includegraphics[width=\textwidth]{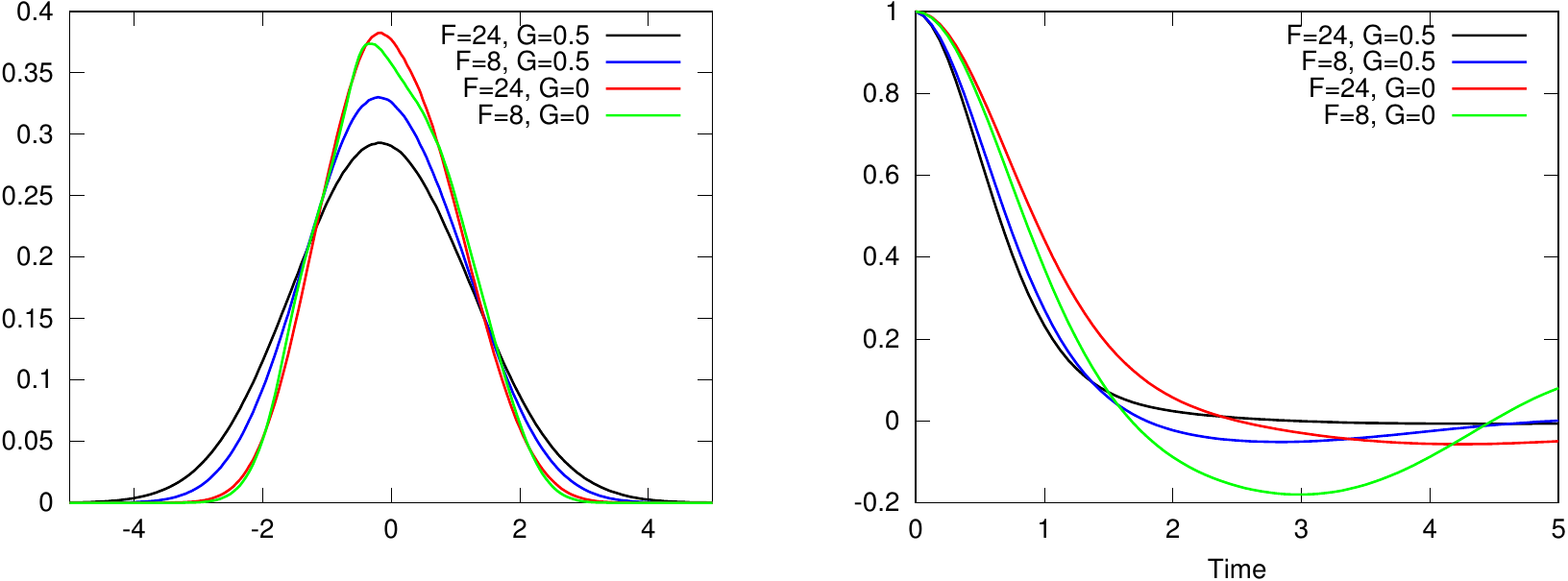}%
\caption{Probability density functions and time autocorrelation
  functions.}%
\label{fig:pdfs_corrs}%
\end{figure}

The computational settings for the numerical simulations were chosen
as follows:
\begin{itemize}
\item Forward integration scheme: Runge-Kutta 4th order
for the deterministic part of the time step, forward Euler for the
stochastic part of the time step;
\item Time discretization step: $\Delta t=1/64$;
\item Time averaging window: $T_{av}=$200000 time units;
\item Spin-up time window (time skipped between the initial condition
  and the beginning of the time averaging window): $T_{skip}=$10000
  time units;
\item Initial condition: each initial state $x_i$, $1\leq i\leq N$, is
  generated at random using normal distribution with zero mean and
  unit standard deviation.
\end{itemize}
In Figure~\ref{fig:pdfs_corrs} we show the histograms of the
probability density functions (PDFs), computed by the standard
bin-counting, as well as the simplest time-lag autocorrelation
functions of the solution with itself, the latter computed numerically
as
\begin{equation}
C(t)=\frac 1{T_{av}} \int_{T_{skip}}^{T_{skip}+T_{av}} x(s)x(t+s)\dif
s,
\end{equation}
where $x(t)$ denotes one of the $N$ variables
of~\eqref{eq:lorenz_rescaled}.  Obviously, due to the translational
invariance of~\eqref{eq:lorenz_rescaled}, both the PDFs and
correlation functions are identical across different variables.
Observe that the PDFs look close to Gaussian, and the time
autocorrelation functions decay rather rapidly within the first five
units of time. There are two reasons why we need to check the decay of
the time autocorrelation functions: first, we need to ensure that the
time averaging window $T_{av}$ is much longer than the decay time
scale of $C(t)$ for the adequate statistical averaging; and, second,
we need to estimate the time scale of development of the response,
since it is directly connected to the time scale of the
autocorrelation functions according to~\eqref{eq:R_1_2}
and~\eqref{eq:R_2_2}.
\begin{table}
\begin{center}%
\begin{tabular}{|c||c|c|}
\hline
 & Skewness & Kurtosis \\
\hline\hline
 $F=24, G=0.5$ & $3.806\cdot 10^{-2}$ & $2.92$ \\
 $F=8, G=0.5$ & $6.42\cdot 10^{-2}$ & $2.81$ \\
 $F=24, G=0$ & $0.1059$ & $2.699$ \\
 $F=8, G=0$ & $9.375\cdot 10^{-2}$ & $2.481$ \\
\hline
\end{tabular}
\end{center}%
\vspace{0.5EM}
\caption{The skewness and kurtosis.}
\label{tab:skewness_kurtosis}
\end{table}

For more precise estimates of how close the PDFs on
Figure~\ref{fig:pdfs_corrs} are to Gaussian, in
Table~\ref{tab:skewness_kurtosis} we show the skewness (third moment)
and kurtosis (fourth moment) of the PDFs, nondimensionalized by the
appropriate powers of the variance. For the purely Gaussian
distribution, the skewness is zero, and the kurtosis is 3. Observe
that, in this respect, the PDFs for the dynamical regimes with greater
$F$ and greater $G$ are closer to the Gaussian, and thus we may expect
generally better performance of the qG-FDT approximations
in~\eqref{eq:R_1_2} and~\eqref{eq:R_2_2} for those regimes.

\subsection{The external perturbations and the response function}

For testing the response theory developed above, we use a rather
simple set-up. We set the stochastic perturbation matrix $\BM H$
entirely to zero, except for its single upper-left corner element,
which is set to 1:
\begin{equation}
\label{eq:H}
\BM H(\BV x)=\left(\begin{array}{cccc} 1 & 0 & \cdots & 0 \\
0 & 0 & \cdots & 0 \\ \vdots & \vdots & \ddots & \vdots \\
0 & 0 & \cdots & 0 \\ \end{array}\right).
\end{equation}
The product $\varepsilon\eta(t)$ is set to a small constant $\eta$ at
zero response time:
\begin{equation}
\label{eq:eta}
\varepsilon\eta(t)=\left\{\begin{array}{c@{\quad}c} 0, & t<0,\\ \eta,
& t\geq 0,\\\end{array}\right.
\end{equation}
where we choose two values for $\eta$: $\eta=0.05$, and $\eta=0.1$.
Thus, the stochastic perturbation is applied to the first variable of
the model, $x_1$, and constitutes a scalar Wiener noise $\eta W_t$.

As far as the choice of the response function $A(\BV x)$ is concerned,
it is obvious that monitoring a single scalar quantity (as presented
in the theory above) is not sufficient to evaluate the detailed impact
of the external stochastic perturbation, even of the simplest type we
chose above, on the model. Thus, we choose to monitor the response to
the stochastic perturbation of each model variable instead. More
precisely, instead of monitoring one response function for the whole
system, we monitor $N$ of those, separately for each model variable:
\begin{equation}
A_i(\BV x)=x_i^2,\qquad 1\leq i\leq N,
\end{equation}
where the square of a variable $x_i$ is chosen (rather than the
variable itself) because it is likely to respond more substantially to
a stochastic perturbation. If we denote the set of all $A_i(\BV x)$ as
the vector $\BV A(\BV x)$, the latter can be expressed concisely as
the Hadamard product of $\BV x$ with itself:
\begin{equation}
\label{eq:response_function}
\BV A(\BV x)=\BV x\circ\BV x.
\end{equation}
Note that in our previous
works~\cite{Abr5,Abr6,Abr7,Abr12,AbrKje,AbrMaj4,AbrMaj5,AbrMaj6,MajAbrGro},
where the deterministic perturbations where studied, the typical
choice of $A(\BV x)$ was $\BV x$ itself. Here, however, we
prefer~\eqref{eq:response_function} because $\BV x$ by itself is
unlikely to respond to a stochastic perturbation in a well-articulated
fashion.

The set-up above allows us to compute the response of all $x_i^2$,
$1\leq i\leq N$, separately, to a small constant stochastic forcing at
the first variable, $x_1$, which is switched on at zero time. Due to
the constant nature of the stochastic perturbation, the response
formulas are simplified as
\begin{equation}
\delta\langle\BV A\rangle(t)=\mathcal R_1(t)\eta,\qquad\mathcal
R_1(t)=\int_0^t\BV R_1(s)\dif s,
\end{equation}
or
\begin{equation}
\delta\langle\BV A\rangle(t)=\mathcal R_2(t)\eta^2,\qquad\mathcal
R_2(t)=\int_0^t\BV R_2(s)\dif s,
\end{equation}
depending on the type of stochastic forcing, where $\BV R_1(s)$ and
$\BV R_2(s)$ are the corresponding vectorized leading order qG-FDT
response operators from~\eqref{eq:R_1_2} and~\eqref{eq:R_2_2} for the
vector response function in~\eqref{eq:response_function}. In what
follows, we display the operators $\mathcal R_1(t)$ or $\mathcal
R_2(t)$ (again, depending on the type of stochastic perturbation)
computed at different times $t=0.5,1,2,4$. We compare these operators
with the directly measured responses via ensemble simulations,
normalized by either $\eta$ (for comparison with $\mathcal R_1$) or
$\eta^2$ (for comparison with $\mathcal R_2$), respectively. The
ensemble simulations are performed with the ensemble size of 20000
members (which were sampled from the same long term trajectory of the
unperturbed system as was used to compute the statistics and the
qG-FDT response), with 1000 realizations of the Wiener process carried
out for each member. Each such ensemble simulation took several hours
on a 16-processor Intel Xeon server, running fully in parallel. In
contrast, each qG-FDT response computation took only few minutes using
a single CPU core.

In addition to the plots of the response at different times, we show
the relative errors and collinearity correlations between the actually
measured response, and the response predicted by the qG-FDT formulas
in~\eqref{eq:R_1_2} or~\eqref{eq:R_2_2}, depending on the type of
noise perturbation. The relative error is defined as the ratio of the
Euclidean norm of the difference between the qG-FDT response and the
normalized measured response, over the Euclidean norm of the qG-FDT
response:
\begin{equation}
\mbox{Relative error}=\begin{cases} \frac{\left\|\frac{\delta\langle
    \BV A\rangle_{\mathrm{measured}}}\eta-\mathcal R_1\right\|}{
  \|\mathcal R_1\|}\quad\mbox{when perturbing the existing noise
  term},\\ \frac{\left\|\frac{\delta\langle \BV A\rangle_{
      \mathrm{measured}}}{\eta^2}-\mathcal R_2 \right\|}{\|\mathcal
  R_2\|}\quad\mbox{when perturbing with a new noise term}.
\end{cases}
\end{equation}
The collinearity correlation is defined as the Euclidean inner product
of the qG-FDT response with one of the measured responses, normalized
by the product of their corresponding Euclidean norms:
\begin{equation}
\mbox{Collinearity correlation}=\frac{\delta\langle\BV A\rangle_{
    \mathrm{measured}} \cdot\mathcal R}{\|\delta\langle\BV A\rangle_{
    \mathrm{measured}}\|\|\mathcal R\|}.
\end{equation}
It is easy to see that the collinearity correlation achieves its
maximum value of 1 if and only if one response is the exact multiple
of the other.

\subsection{Perturbing the existing noise term}

\begin{figure}
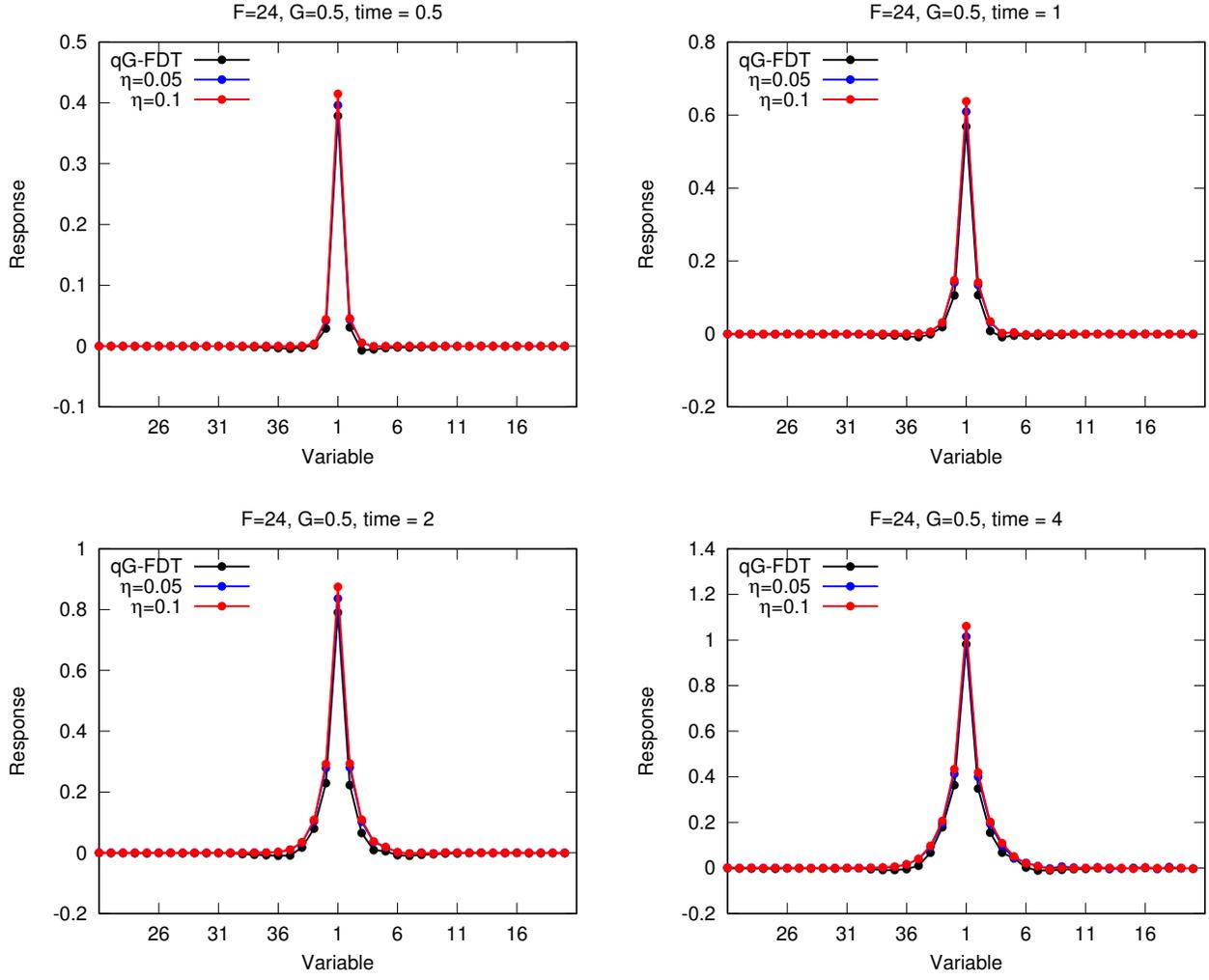

\includegraphics[width=\textwidth]{{{FIG_response_L24_n0.5}}}%
\caption{Response to the existing noise perturbation, $F=24$,
  $G=0.5$.}%
\label{fig:response_L24_n0.5}
\end{figure}

In this section we study the response of two dynamical regimes of the
stochastically forced ($G=0.5$) rescaled Lorenz 96 model
in~\eqref{eq:lorenz_rescaled}. The first regime corresponds to the
constant forcing $F=24$, and is, according to
Figure~\ref{fig:pdfs_corrs} and Table~\ref{tab:skewness_kurtosis}, the
closest to the Gaussian regime of all examined in
Section~\ref{sec:stats}. In Figure~\ref{fig:response_L24_n0.5} we show
the response of the function~\eqref{eq:response_function} to the
perturbation of the existing stochastic matrix $\BM G=0.5\BM I$ by the
perturbations described above in~\eqref{eq:H}--\eqref{eq:eta}, with
$\eta$ set to 0.05 and 0.1. With help of the periodicity
of~\eqref{eq:lorenz_rescaled}, the response variables in
Figure~\ref{fig:response_L24_n0.5}, as well as all subsequent figures,
are displayed so that the variable $x_1$ (on which the perturbation
occurs) is at the center of the plot, with $x_2$ immediately to the
right, and $x_N$ to the left.

Observe that the precision of the qG-FDT response prediction in this
case is truly striking -- there is hardly any visual difference
between the qG-FDT prediction and the directly measured responses for
both $\eta=0.05$ and $\eta=0.1$.
Additionally, in Table~\ref{tab:errors_L24_n0.5} we show the relative
errors and collinearity correlations for the qG-FDT prediction and two
directly measured responses for the regime with $F=24$ and
$G=0.5$. Observe that the relative errors are about 10-15\% for all
displayed response times, and their collinearity exceeds 99\%.

In Figure~\ref{fig:response_L8_n0.5} we show the results for the
second regime where we perturb the existing stochastic forcing. In
this regime, the stochastic diffusion matrix is set to the same value
$\BM G=0.5\BM I$ as before, by the constant deterministic forcing $F$
is set to $F=24$. According to Figure~\ref{fig:pdfs_corrs} and
Table~\ref{tab:skewness_kurtosis}, this regime is the second closest
to the Gaussian, and it is clearly manifested in the difference
between the qG-FDT prediction and the directly measured response with
$\eta=0.05$ and $\eta=0.1$, which are shown in
Figure~\ref{fig:response_L8_n0.5}. Namely, this time there is a
visible discrepancy between the qG-FDT response prediction and both
directly measured responses in the second variable to the right from
the one where the perturbation is applied.
Table~\ref{tab:errors_L8_n0.5} reinforces this visual perception,
indicating relative errors between 15\% and 26\%. Surprisingly, the
collinearity correlations between the qG-FDT prediction and the
directly measured responses do not suffer much, still remaining about
97-98\% of the maximum possible.
\begin{table}
\begin{center}%
\begin{tabular}{|c|c|}
\hline
 Rel. Errors, $F=24, G=0.5$ & Coll. Corrs, $F=24, G=0.5$ \\
\hline
\begin{tabular}{c||c|c}
 & $\eta=0.05$ & $\eta=0.1$ \\
\hline\hline
$t=0.5$ & $7.82\cdot 10^{-2}$ & $0.1178$ \\
$t=1$ & $0.1181$ & $0.1612$ \\
$t=2$ & $0.1274$ & $0.1692$ \\
$t=4$ & $9.826\cdot 10^{-2}$ & $0.1383$ \\
\end{tabular}
&
\begin{tabular}{c||c|c}
 & $\eta=0.05$ & $\eta=0.1$ \\
\hline\hline
$t=0.5$ & $0.9984$ & $0.9984$ \\
$t=1$ & $0.9972$ & $0.9972$ \\
$t=2$ & $0.9964$ & $0.9964$ \\
$t=4$ & $0.9975$ & $0.9974$ \\
\end{tabular}\\
\hline
\end{tabular}
\end{center}%
\vspace{0.5EM}
\caption{Relative errors and collinearity correlations in the response
  to the existing noise perturbation, $F=24$, $G=0.5$.}
\label{tab:errors_L24_n0.5}
\end{table}
\begin{figure}
\includegraphics[width=\textwidth]{{{FIG_response_L8_n0.5}}}%
\caption{Response to the existing noise perturbation, $F=8$, $G=0.5$.}%
\label{fig:response_L8_n0.5}
\end{figure}
\begin{table}
\begin{center}%
\begin{tabular}{|c|c|}
\hline
 Rel. Errors, $F=8, G=0.5$ & Coll. Corrs, $F=8, G=0.5$ \\
\hline
\begin{tabular}{c||c|c}
 & $\eta=0.05$ & $\eta=0.1$ \\
\hline\hline
$t=0.5$ & $0.1464$ & $0.1643$ \\
$t=1$ & $0.2116$ & $0.2354$ \\
$t=2$ & $0.2385$ & $0.2614$ \\
$t=4$ & $0.2046$ & $0.224$ \\
\end{tabular}
&
\begin{tabular}{c||c|c}
 & $\eta=0.05$ & $\eta=0.1$ \\
\hline\hline
$t=0.5$ & $0.9898$ & $0.9898$ \\
$t=1$ & $0.9807$ & $0.9807$ \\
$t=2$ & $0.9748$ & $0.9746$ \\
$t=4$ & $0.98$ & $0.9796$ \\
\end{tabular}\\
\hline
\end{tabular}
\end{center}%
\vspace{0.5EM}
\caption{Relative errors and collinearity correlations in the response
  to the existing noise perturbation, $F=8$, $G=0.5$.}
\label{tab:errors_L8_n0.5}
\end{table}

\subsection{Perturbing with a new noise}

Here we show the results of numerical simulations where the external
stochastic perturbation is introduced into the system via a separate
noise realization. In this situation, we can study both the fully
deterministic and the stochastically forced dynamics of the rescaled
Lorenz 96 model in~\eqref{eq:lorenz_rescaled}; together with two
different values of the forcing $F$, this constitutes four different
combinations of parameters: ($F=24$, $G=0$), ($F=8$, $G=0$), ($F=24$,
$G=0.5$), and ($F=8$, $G=0.5$).

As we mentioned before, the qG-FDT approximation in~\eqref{eq:R_2_2}
should not formally be valid for the deterministic unperturbed
dynamics with $G=0$, since there is no guarantee that the invariant
measure $\mu$ of the unperturbed dynamics is even continuous with
respect to the Lebesgue measure, let alone possesses a Gaussian
density. However, practice showed in the past with deterministic
perturbations~\cite{AbrMaj4,AbrMaj5,AbrMaj6,MajAbrGro,AbrKje} that the
quasi-Gaussian formula for the deterministic invariant measure can in
fact be quite a reasonable approximation, especially if the
unperturbed dynamics is strongly chaotic and rapidly mixing. This is
what we confirm here for stochastic external perturbations as well.

To model the ideal response in the stochastically forced unperturbed
regimes with $G=0.5$, we apply an independent realization of the
unperturbed stochastic forcing with the diffusion matrix $\BM G=0.5\BM
I$ to each of the 20000 ensemble members, while modeling the
stochastic perturbation via another 1000 independent realizations of
the Wiener noise for each ensemble simulation as before. This is done
to retain the same computational expense as for the other studied
cases, which, of course, leads to statistical undersampling of the
expectation over the noise realizations. Indeed, in the case of
stochastic perturbations independent from the unperturbed noise, the
expectation must be computed over the comparable number of
realizations separately for the unperturbed noise and for the
stochastic perturbation to retain comparable averaging
fidelity. Still, we find that even in this simplified setup the qG-FDT
approximation shows good agreement with the measured response, at
least for sufficiently short response times.

\begin{figure}
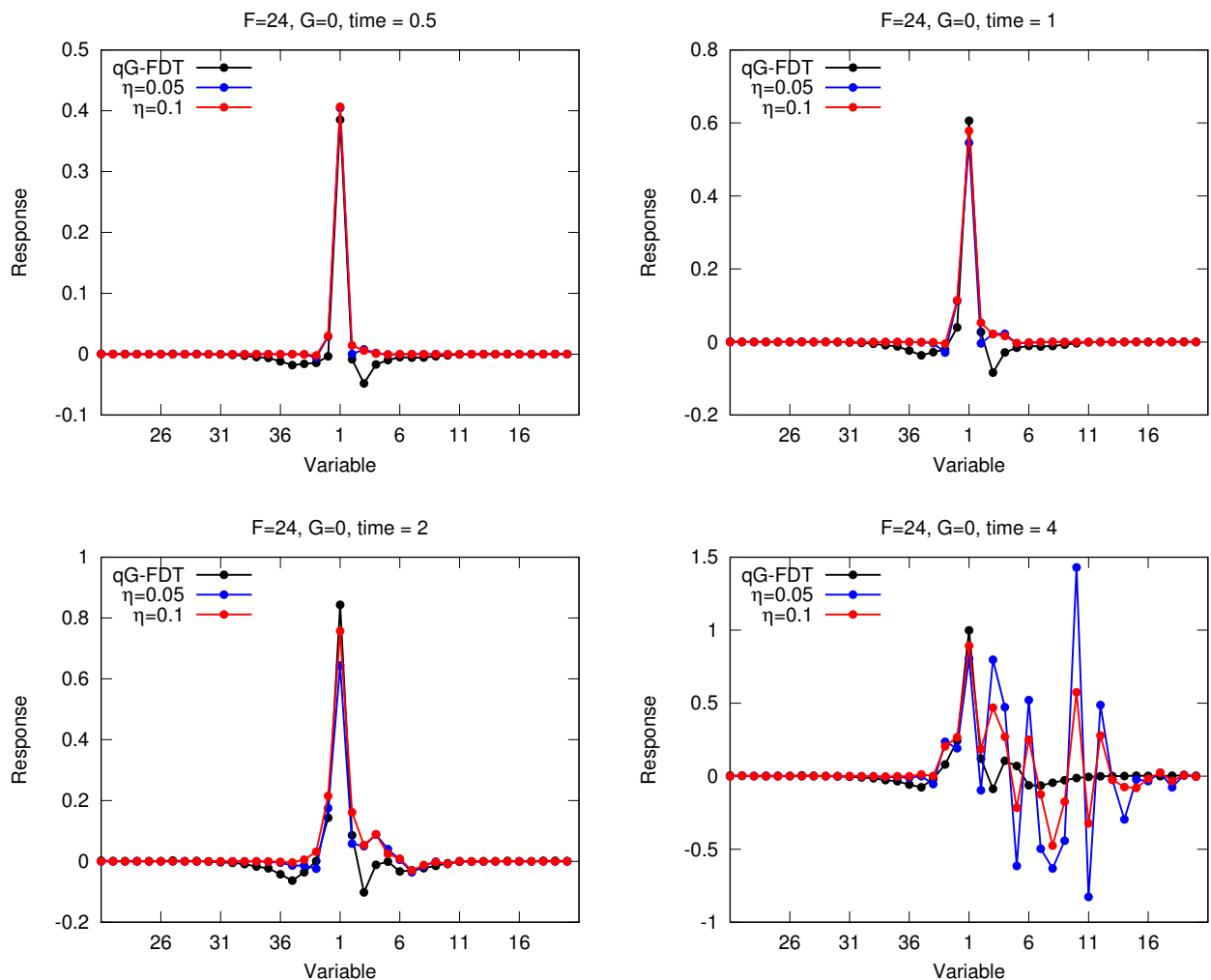

\includegraphics[width=\textwidth]{{{FIG_response_L24_n0}}}%
\caption{Response to the perturbation with a new noise, $F=24$, $G=0$.}%
\label{fig:response_L24_n0}
\end{figure}
\begin{table}
\begin{center}%
\begin{tabular}{|c|c|}
\hline
 Rel. Errors, $F=24, G=0$ & Coll. Corrs, $F=24, G=0$ \\
\hline
\begin{tabular}{c||c|c}
 & $\eta=0.05$ & $\eta=0.1$ \\
\hline\hline
$t=0.5$ & $0.199$ & $0.206$ \\
$t=1$ & $0.2657$ & $0.25$ \\
$t=2$ & $0.3325$ & $0.2906$ \\
$t=4$ & $2.241$ & $1.081$ \\
\end{tabular}
&
\begin{tabular}{c||c|c}
 & $\eta=0.05$ & $\eta=0.1$ \\
\hline\hline
$t=0.5$ & $0.9817$ & $0.9807$ \\
$t=1$ & $0.966$ & $0.9683$ \\
$t=2$ & $0.96$ & $0.9572$ \\
$t=4$ & $0.3126$ & $0.6303$ \\
\end{tabular}\\
\hline
\end{tabular}
\end{center}%
\vspace{0.5EM}
\caption{Relative errors and collinearity correlations in the response
  to the perturbation with a new noise, $F=24$, $G=0$.}
\label{tab:errors_L24_n0}
\end{table}
\begin{figure}
\includegraphics[width=\textwidth]{{{FIG_response_L8_n0}}}%
\caption{Response to the perturbation with a new noise, $F=8$, $G=0$.}%
\label{fig:response_L8_n0}
\end{figure}
\begin{table}
\begin{center}%
\begin{tabular}{|c|c|}
\hline
 Rel. Errors, $F=8, G=0$ & Coll. Corrs, $F=8, G=0$ \\
\hline
\begin{tabular}{c||c|c}
 & $\eta=0.05$ & $\eta=0.1$ \\
\hline\hline
$t=0.5$ & $0.3602$ & $0.3672$ \\
$t=1$ & $0.4952$ & $0.443$ \\
$t=2$ & $0.5291$ & $0.4581$ \\
$t=4$ & $1.112$ & $0.7775$ \\
\end{tabular}
&
\begin{tabular}{c||c|c}
 & $\eta=0.05$ & $\eta=0.1$ \\
\hline\hline
$t=0.5$ & $0.9358$ & $0.9347$ \\
$t=1$ & $0.869$ & $0.8967$ \\
$t=2$ & $0.8719$ & $0.8896$ \\
$t=4$ & $0.5373$ & $0.7163$ \\
\end{tabular}\\
\hline
\end{tabular}
\end{center}%
\vspace{0.5EM}
\caption{Relative errors and collinearity correlations in the response
  to the perturbation with a new noise, $F=8$, $G=0$.}
\label{tab:errors_L8_n0}
\end{table}
In Figure~\ref{fig:response_L24_n0} and Table~\ref{tab:errors_L24_n0}
we show the qG-FDT prediction together with the directly measured
response with perturbations $\eta=0.05$ and $\eta=0.1$, for the
dynamical regime of~\eqref{eq:lorenz_rescaled} with $F=24$ and $G=0$
(fully deterministic). Observe that, in comparison with the previously
studied regimes, the quality of the qG-FDT prediction tends to
deteriorate rather substantially, as the relative errors increase to
20-33\%, and the collinearity correlations drop to 95-98\% for the
response times $t\leq 2$. However, what we can also observe is the
large discrepancy between the two directly measured responses for
different perturbation magnitudes, which develops at $t=4$. Note that
no such discrepancy was observed for the stochastic unperturbed
dynamics -- both directly measured responses, for $\eta=0.05$ and
$\eta=0.1$, were virtually identical in
Figures~\ref{fig:response_L24_n0.5} and~\ref{fig:response_L8_n0.5}.
This could be a manifestation of the developing structural instability
of the attractor of the system as the dynamics transition from purely
deterministic to stochastic.

In Figure~\ref{fig:response_L8_n0} and Table~\ref{tab:errors_L8_n0} we
show the results for the regime with $F=8$, and all other parameters
set as above. Here, observe that the quality of the qG-FDT prediction
deteriorates even further, as the relative errors increase to 36-53\%,
and the collinearity correlations drop to 93-87\% for the response
times $t\leq 2$. This is to be expected, however, since this regime is
the farthest from the Gaussian, according to
Figure~\ref{fig:pdfs_corrs} and Table~\ref{tab:skewness_kurtosis}. The
discrepancy between the directly measured responses for perturbations
of different magnitudes, which could be attributed to the developing
structural instability of the attractor, also manifests itself at time
$t=4$, as for the previously studied deterministic regime with $F=24$.

\begin{figure}
\includegraphics[width=\textwidth]{{{FIG_response_L24_new_noise_n0.5}}}%
\caption{Response to the perturbation with a new noise, $F=24$, $G=0.5$.}%
\label{fig:response_L24_new_noise_n0.5}
\end{figure}
\begin{table}
\begin{center}%
\begin{tabular}{|c|c|}
\hline
 Rel. Errors, $F=24, G=0.5$ & Coll. Corrs, $F=24, G=0.5$ \\
\hline
\begin{tabular}{c||c|c}
 & $\eta=0.05$ & $\eta=0.1$ \\
\hline\hline
$t=0.5$ & $0.1343$ & $8.118 \cdot 10^{-2}$ \\
$t=1$ & $0.1674$ & $8.757 \cdot 10^{-2}$ \\
$t=2$ & $0.4152$ & $0.3859$ \\
$t=4$ & $5.17$ & $1.506$ \\
\end{tabular}
&
\begin{tabular}{c||c|c}
 & $\eta=0.05$ & $\eta=0.1$ \\
\hline\hline
$t=0.5$ & $0.9975$ & $0.9987$ \\
$t=1$ & $0.9859$ & $0.9962$ \\
$t=2$ & $0.9099$ & $0.9225$ \\
$t=4$ & $0.222$ & $0.585$ \\
\end{tabular}\\
\hline
\end{tabular}
\end{center}%
\vspace{0.5EM}
\caption{Relative errors and collinearity correlations in the response
  to the perturbation with a new noise, $F=24$, $G=0.5$.}
\label{tab:errors_L24_new_noise_n0.5}
\end{table}
\begin{figure}
\includegraphics[width=\textwidth]{{{FIG_response_L8_new_noise_n0.5}}}%
\caption{Response to the perturbation with a new noise, $F=8$, $G=0.5$.}%
\label{fig:response_L8_new_noise_n0.5}
\end{figure}
\begin{table}
\begin{center}%
\begin{tabular}{|c|c|}
\hline
 Rel. Errors, $F=8, G=0.5$ & Coll. Corrs, $F=8, G=0.5$ \\
\hline
\begin{tabular}{c||c|c}
 & $\eta=0.05$ & $\eta=0.1$ \\
\hline\hline
$t=0.5$ & $0.1523$ & $0.1339$ \\
$t=1$ & $0.3305$ & $0.217$ \\
$t=2$ & $0.4922$ & $0.2864$ \\
$t=4$ & $2.381$ & $1.097$ \\
\end{tabular}
&
\begin{tabular}{c||c|c}
 & $\eta=0.05$ & $\eta=0.1$ \\
\hline\hline
$t=0.5$ & $0.9891$ & $0.9911$ \\
$t=1$ & $0.9467$ & $0.9795$ \\
$t=2$ & $0.8971$ & $0.9673$ \\
$t=4$ & $0.3568$ & $0.7128$ \\
\end{tabular}\\
\hline
\end{tabular}
\end{center}%
\vspace{0.5EM}
\caption{Relative errors and collinearity correlations in the response
  to the perturbation with a new noise, $F=8$, $G=0.5$.}
\label{tab:errors_L8_new_noise_n0.5}
\end{table}
In Figure~\ref{fig:response_L24_new_noise_n0.5} and
Table~\ref{tab:errors_L24_new_noise_n0.5} we show the qG-FDT
prediction together with the directly measured response with
perturbations $\eta=0.05$ and $\eta=0.1$, for the dynamical regime
of~\eqref{eq:lorenz_rescaled} with $F=24$ and $G=0.5$, where the
stochastic perturbation is independent of the unperturbed noise. It is
interesting that, in comparison with the previously studied regimes,
the quality of the qG-FDT prediction tends to be good for short
response times, $t=0.5$ and $t=1$, with relative errors about 8-16\%,
and collinearity correlations about 98-99\%. However, for longer
response times the errors appear to grow faster than in any of the
previously considered cases.

Just as for the deterministic unperturbed dynamics above, here we can
also observe the large discrepancy between the two directly measured
responses for different perturbation magnitudes, which develops at
$t=2$. This is likely the manifestation of statistical undersampling
mentioned above, since the size of the statistical ensemble is
unchanged even though an additional independent stochastic forcing is
introduced into the dynamics.

In Figure~\ref{fig:response_L8_new_noise_n0.5} and
Table~\ref{tab:errors_L8_new_noise_n0.5} we show the results for the
regime with $F=8$, $G=0.5$, and a statistically independent stochastic
perturbation. Here, the observations appear to be consistent with the
previous results. Namely, the relative errors (15-33\%) for the
initial times $t=0.5$ and $t=1$ are worse than those for the
previously considered regime, $F=24$, $G=0.5$, likely due to the fact
that the qG-FDT approximation is worse for the present regime due to
stronger non-Gaussianity of the unperturbed regime. However, the
errors for longer times, $t=2$ and $t=4$, are smaller than those for
the regime with $F=24$, again, likely due to the fact that the
statistical undersampling does not manifest as strongly in a less
chaotic regime. The collinearity correlations appear to follow the
same trend as the relative errors.

\section{Summary}
\label{sec:summary}

In this work we develop a fluctuation-response theory and test a
computational algorithm for the leading order response of chaotic and
stochastic dynamical systems to stochastic perturbations. The key
property of this approach is that it allows to estimate the average
response to an external stochastic perturbation from a certain
combination of the time-lagged averages of the unperturbed system. For
dynamical systems, which are already stochastic, we consider two
cases: first, where the existing stochastic term is perturbed; and,
second, where a new stochastic perturbation is introduced, which
correspondingly leads to different leading order average response
formulas. We also show that, under appropriate assumptions, the
resulting formulas for leading order response to a stochastic
perturbation for the deterministic and stochastic unperturbed dynamics
are equivalent. For the practical computation of the leading order
response approximation, we derive the approximate quasi-Gaussian
response formulas, where the probability density of the unperturbed
statistical state is assumed to be Gaussian. We numerically
investigate the validity of the quasi-Gaussian response formulas for
stochastic perturbations of both deterministic and stochastically
forced Lorenz 96 system. We find that the quasi-Gaussian response
formulas appear to more effective for the regimes where the
unperturbed dynamics is already stochastic. Additionally, for the
stochastic perturbations of the deterministic Lorenz 96 model, we
observe what seems to be a manifestation of structural instability of
the system's attractor under a stochastic perturbation.

\ack The work was supported by the Office of Naval Research grant
N00014-15-1-2036.

\end{document}